\documentclass[prb,aps,twocolumn,showpacs,superscriptaddress]{revtex4}
\usepackage[dvips]{graphicx}
\usepackage{latexsym}
\usepackage{amsmath}
\begin{document}

\title{Higher angular momentum Kondo liquids}

\author{Pouyan Ghaemi}
\affiliation{ Department of Physics, Massachusetts Institute of
Technology, Cambridge, Massachusetts 02139}

\author{T. Senthil}
\affiliation{ Department of Physics, Massachusetts Institute of
Technology, Cambridge, Massachusetts 02139}
\affiliation{Center for
Condensed Matter Theory, Indian Institute of Science, Bangalore,
India 560012}

\date{\today}
\begin{abstract}
Conventional heavy Fermi liquid phases of Kondo lattices  involve
the formation of a ``Kondo singlet" between the local moments and
the conduction electrons. This Kondo singlet is usually taken to be
in an internal s-wave angular momentum state. Here we explore the
possibility of phases where the Kondo singlet has internal angular
momentum that is d-wave.  Such states are readily accessed in a
slave boson mean field formulation, and are energetically favorable
when the Kondo interaction is between a local moment and an electron
at a nearest neighbor site. The properties of the d-wave Kondo
liquid are studied. Effective mass and quasiparticle residue show
large angle dependence on the Fermi surface. Remarkably in certain
cases, the quasiparticle residue goes to zero at isolated points (in
two dimensions) on the Fermi surface. The excitations at these
points then include a free fractionalized spinon. We also point out
the possibility of quantum Hall phenomena in two dimensional Kondo
{\em insulators}, if the Kondo singlet has complex internal angular
momentum. We suggest that such d-wave Kondo pairing may provide a
useful route to thinking about correlated Fermi liquids with strong
anisotropy along the Fermi surface.

\end{abstract}
\newcommand{\fig}[2]{\includegraphics[width=#1]{#2}}
\newcommand{\be}{\begin{equation}}
\newcommand{\ee}{\end{equation}}
\maketitle
\section{Introduction}
In many rare earth alloys, a periodic lattice of localized
spin-$1/2$ magnetic moments is coupled through Kondo
exchange\cite{kondo} to a separate band of conduction
electrons\cite{sdon}. An interesting low temperature metallic phase
often develops at low temperature where the local moments are
absorbed into the Fermi sea by a lattice analog of the Kondo
effect. The resulting metallic phase is a Fermi
liquid albeit with strongly renormalized parameters the most
dramatic of which is a large quasiparticle mass enhancement (leading
to the name `heavy fermi liquid'). Concomittantly the quasiparticle
residue at the Fermi surface is very small (though non-zero).
Perhaps most strikingly the Fermi surface is large in the sense that
its volume satisfies Luttinger's theorem\cite{lutt} only if the
local moments are included in the count of the electron density.
This and other universal properties of this Fermi liquid are
usefully understood in a strong Kondo coupling picture where each
conduction electron is trapped into a spin singlet state with a
local moment - see Ref. \onlinecite{nozieres}.

The singlet `molecule' formed out of the local moment and conduction
electron is usually taken to be in an s-wave state with zero
internal angular momentum. In this paper we explore metallic states
where this singlet has non-zero internal angular momentum. We show
that this results in Fermi liquid states that have many unusual and
interesting properties. For instance such states naturally have
large anisotropies in the effective mass, quasiparticle residue and
other properties on moving around the Fermi surface. Under certain
conditions it is even possible for the quasiparticle residue to
vanish at isolated points of the Fermi surface. The excitations at
such points do not have electron quantum numbers but can be
understood as a neutral fermionic spin-$1/2$ `spinon'. Remarkably
such a spinon excitation emerges without any associated gauge
interaction (unlike the other many familiar examples\cite{ReSaSpN,Wen91,z2long}). In
the special case where the conduction band is half-filled the usual
s-wave Kondo singlet formation leads to an insulating state (dubbed
the Kondo insulator). If the internal angular momentum is non-zero,
interesting varieties of Kondo insulators become possible. For
instance we show that in two dimensions with a $d_{x^2 - y^2} + i
d_{xy}$ singlet, the Kondo insulator has a non-trivial quantized
electrical Hall conductivity.

A partial motivation for studying such higher angular momentum
singlet formation between local moments and conduction electrons
comes from observations on cuprate high-$T_c$ materials. We should
immediately emphasize though that the two band Kondo-lattice models
are not directly expected to be an appropriate microscopic
description of the cuprates. Nevertheless it seems worthwhile to
point out some resemblances between the properties of the metallic
states described in this paper and those seen in the cuprates.
Extending the ideas of this paper to one-band models appropriate to
the cuprates may be an interesting direction for future work. For
some attempts along similar directions see Ref. \onlinecite{tiago}.
The optimally doped cuprates in the normal state are metallic and
have a full Fermi surface satisfying the usual Luttinger
theorem\cite{prev}. Despite this however a true sharp electronic
quasiparticle peak does not seem to exist. The electron spectrum -
measured through photoemission experiments - shows significant
anisotropy on moving around the Fermi surface\cite{dicho}. In
particular the energy distribution curve (EDC) is narrower along the
nodal direction as compared to the antinodal direction. This kind of
anisotropy in the EDC persists into the underdoped side when a
pseudogap opens along the antinodal direction.  It also persists
into the superconducting state obtained by cooling. Collectively
these phenomena have been dubbed the `nodal-antinodal dichotomy':
the nodal excitations are much more quasiparticle-like than the
antinodal ones.

The general message from these results on the cuprates for the study
of correlated metals is that correlation effects may not uniformly affect all parts of the Fermi surface in a metal.
Some portions of the Fermi surface may be more susceptible to correlation effects than others. This will lead to
strong correlation induced anisotropy in physical properties around the Fermi surface. Theoretically study of such effects is difficult because electron correlations are
typically most easily handled in real space which does not readily distinguish between different parts of the Fermi surface. Interesting recent
numerical calculations\cite{kotliar} based on cluster extensions of dynamical mean field theory on single band Hubbard models in two dimensions
have found such momentum space differentiation in the electronic properties.
The ideas of the present paper may be seen as a step toward incorporating  momentum space differentiation into simple
analytic treatments of strong correlation problems.
Anisotropic spectra
at different regions of the Fermi surface occur naturally in the
metallic states described in the two band Kondo lattice models studied in this
paper.

Theoretically higher angular momentum Kondo liquids are
conveniently accessed through the slave boson mean field theory\cite{read,millee}
developed to describe the usual heavy fermi liquid state. In this
theory the local moment $\vec S_r$ at site $r$ is represented in
terms of neutral spin-$1/2$ fermions $f_{r\alpha}$ ($\alpha =
\uparrow, \downarrow$)through
\begin{equation}
\vec S_r = \frac{1}{2}f^{\dagger}_r \vec \sigma f_r
\end{equation}
The Kondo singlet formation is then signalled by the development of
a non-zero expectation value of the singlet hybridization amplitude
$\langle c^{\dagger}_rf_r \rangle$ between the $f$-fermions and
conduction electrons $c_r$. This immediately suggests that Kondo
singlets of higher angular momentum can be described by
hybridization amplitudes between $f_r$ at site $r$ and $c_{r'}$ at a
different site $r'$. Specifically
\begin{equation}
b_{rr'} = \langle c^{\dagger}_r f_{r'} \rangle
\end{equation}
may be viewed as the wave function of the Kondo singlet. Thus by
choosing the internal angular momentum associated with rotations of
the relative coordinate $\vec{r}\ ' - \vec{r}$ appropriately Kondo
liquids with higher angular momentum may be constructed. In  this
paper we will exploit this strategy to construct mean field
descriptions of various such Kondo liquid states. Within the mean
field description, higher angular momentum Kondo liquids correspond
to a particular form of momentum dependence of the Kondo
hybridization amplitude. Momentum dependent hybridization amplitudes
have been previously considered in Ref. \onlinecite{morcol}. Recent
optical transport experiments on the 1-1-5 materials have also been
interpreted in terms of momentum dependent hybridization
amplitudes\cite{burch}\cite{burchb}.

We begin in Section \ref{sbrev} with a brief review of the usual
slave boson mean field theory for a heavy fermi liquid\cite{cole}.
This mean field theory describes an on-site ``s-wave" Kondo singlet.
We then consider in Section \ref{dklmft} a Kondo lattice Hamiltonian
where each local moment interacts through Kondo exchange with a
conduction electron at a neighboring site. Within mean field theory
this Hamiltonian is shown to stabilize a $d$-wave Kondo liquid. The
properties of this state are then studied in Section \ref{dpro}.
Next we modify the Kondo lattice model of Section \ref{dklmft} by
introducing explicit Heisenberg exchange between nearest neighbor
moments. A mean field treatment of this model (Section \ref{dklh})
leads to a metallic state where the quasiparticle residue vanishes
at isolated points of the Fermi surface. As mentioned above this
remarkable state has spinon excitations at these isolated points
without any associated gauge interactions. Next in Section
\ref{kiqh} we consider two dimensional Kondo insulators where the
Kondo singlet has complex internal angular momentum of the form
$d_{x^2-y^2}  + i d_{xy}$. We calculate the Hall conductivity and
show that it has a non-trivial quantized value. Thus this provides
an interesting example of a quantum Hall state in a Kondo lattice.
Section \ref{disc} has some conclusions. various details are in the
Appendices.

\section{Review of mean-field theory for Kondo liquid: \lowercase{s}-Wave singlet}
\label{sbrev}

Consider the Kondo lattice model describing a lattice of spin-$1/2$
local moments $\vec S_i$ and conduction electrons $c_{i\alpha}$
$(\alpha = \uparrow, \downarrow)$ coupled through Kondo
exchange\cite{kondo}.
\begin{equation}\label{skondo}
H=-\sum_{\langle ij \rangle} t c^\dagger_i c_j
+\frac{J_K}{2}\sum_{i} \vec{S}_i.c^{\dagger}_i \vec \sigma c_i
\end{equation}
The total number of
conduction electrons per site is taken to be some fixed value $n_c$.
We also specialize to two dimensional systems though extension to three dimension is straightforward. A useful mean
field treatment is obtained using a fermionic representation of the
local moment spin:
\begin{equation}
\vec{S}_i=f^\dag_i\frac{\vec{\sigma}}{2}f_i
\end{equation}
together with the constraint $f^{\dagger}_i f_i = 1$ at each site.
After some algebra the interaction term ($J_K$) reduces to:
\begin{equation}\nonumber
\sum_i f_i^\dag c_i c^\dag_i f_i+ constant
\end{equation}
In the corresponding partition function, this takes the form
\begin{equation}
e^{\int d\tau \sum_i \bar{f}_i c_i \bar{c}_i f_i}
\end{equation}
 Using a Hubbard-Stratonovich transformation, this can be decoupled as
\begin{equation}\label{hubs}
\int [DV]\  \exp\left(\sum_i \frac{|V(i,\tau)|^2}{4 J_k}+(V(i,\tau)
\bar{f}_i c_i+V^*(i,\tau) \bar{c}_i f_i)\right)
\end{equation}
Exact integration over $V(i,\tau)$ will produce the original
interaction term. But in mean-field approximation, we do the
integral using saddle-point method. In addition the constraint
$f^{\dagger}_i f_i = 1$ will be implemented on average. We will
choose the mean-field solution with $V(i,\tau)=V$ which leads to the
following mean-field Hamiltonian and self consistency equation for
$V$:
\begin{eqnarray}
H&=& - \sum_{\langle ij \rangle} t  c^\dagger_i c_j+ h.c \nonumber \\
& & + \sum_i \mu_f
f^\dag_i f_i+V\sum_i (f^\dag_i c_i+c^\dag_i f_i)\\
V&=& \frac{J_k}{2N} \sum_i \langle f^\dag_i c_i+h.c. \rangle \\
\langle f^{\dagger}_i f_i \rangle & = & 1
\end{eqnarray}
We have introduced a chemical potential term for the $f$-fermions
which serves to set their average number per site to be one. In addition the Fermi energy $E_F$ of the hybridized
quasiparticle must be determined by requiring that there are a total of $1+n_c$ fermions per site. Only states with energy less than $E_F$
are filled in the ground state.
 This mean-field Hamiltonian is quadratic and can be diagonalized. To do so, first we write the
 Hamiltonian in the Fourier space:
 \begin{equation}\label{Hk}
H=\sum_{k} \epsilon_\textbf{k} c^\dag_\textbf{k}
c_\textbf{k}+\sum_\textbf{k} \mu_f f^\dag_\textbf{k}
f_{\textbf{k}}+V\sum_\textbf{k} (f^\dag_\textbf{k}
c_\textbf{k}+c^\dag_\textbf{k} f_\textbf{k})
\end{equation}
with $\epsilon_\textbf{k}=-2t\ (\cos(k_x)+\cos(k_y))$. Now
it is easy to derive quasi-particle dispersion relation for two
bands:
\begin{equation} \label{spectrum}
E^\pm_{\textbf{k}}=\frac{\epsilon_\textbf{k}+\mu_f}{2}\pm\sqrt{(\frac{\epsilon_\textbf{k}-\mu_f}{2})^2+V^2}
\end{equation}
 With this in hand, we can solve the  self consistency conditions and calculate physical properties.  For
 instance the density of states
 at the Fermi energy is exponentially large in $1/J_K$ leading to
 the large mass enhancement at small $J_K$.

\section{\lowercase{d}-wave Kondo liquid}
\label{dklmft}
 In the last section we considered the case where the
singlets between the local moments and conduction electrons are
formed with zero internal angular momentum (i.e. $\langle f^\dag_i
c_i + h.c. \rangle$ is constant). When we assume that the singlet is
formed between the local moments and conduction electron at the same
site (i.e. singlet is local), the s-wave is the only possibility for
internal state of the singlet. In this section we consider modifying
the Kondo lattice Hamiltonian so that Kondo singlets with non-zero
angular momentum are favored. To that end we consider a generalized
Kondo Hamiltonian:
\begin{equation}\label{dkondo}
H=- \sum_{\langle ij \rangle} t c^\dag_i c_j + \sum_{ij} J_{ij} \vec{S}_i.\vec{s}_j
\end{equation}
 where the Kondo exchange term is not limited to the local moments and
 conduction electrons at the same site. The simplest
 case is for $J_{ij}$ to be non-zero only when
 $i$ and $j$ denote the nearest neighbor sites:
\begin{eqnarray}
J_{ij} & = & J_K ~~~~(i,j)~\text{nearest neighbor} \\
J_{ij} & = & 0 ~~~~\text{otherwise}
\end{eqnarray}
 With this choice and after
 some algebra, similar to what we did in the last section, the
 Hamiltonian reduces to the following form:
\begin{equation}
H=- \sum_{\langle ij \rangle} t c^\dag_i c_j + \sum_{\langle ij \rangle} J_{ij} \ f^\dag_i c_j c^\dag_j f_i
\end{equation}
Now proceed as before by decoupling the interacting part of the
action using an auxiliary field, which this time lives on the bonds
of the lattice, instead of the sites:
\begin{equation}\label{hubd}\begin{split}
\int & [DV]\ \\ & e^{-\left(\int d\tau \sum_{\langle i,j \rangle}
\frac{|V(i,j,\tau)|^2}{4 J_k}  +(V(i,j,\tau) \bar{f}_i
c_j+V^*(i,j,\tau) \bar{c}_j f_i)\right)}
\end{split}\end{equation}
As before, to get the mean-field Hamiltonian, we'll do the
integration over $V$ using saddle point approximation (in addition
to imposing $\langle f^{\dagger}_i f_i \rangle = 1$ on average).
There are different saddle point solutions (they are basically
different local minima in functional space). We will consider two
solutions named as s-wave and d-wave (the reason for choosing these
names will become clear later). In s-wave solution we consider
$V(i,j,\tau)=V$ on all bonds and in d-wave, we consider
$V(i,j,\tau)=V$ on $x$ bonds and $V(i,j,\tau)=-V$ on $y$ bonds:
\begin{equation}
\label{Hammf}
\begin{split} H_{\left(
\begin{array}{ccc}
s \\
d \end{array} \right)}=&\sum_{- \langle ij \rangle} t c^\dag_i c_j
 +\sum_i \mu_f f^\dag_i f_i+\\&\sum_{i,x,y}
V [(f_i^\dag c_{i+x} +c^\dag_{i+x} f_i)\pm (f_i^\dag c_{i+y}
+c^\dag_{i+y} f_i)]
\end{split}
\end{equation}

It is obvious from this form that these correspond to s-wave and
d-wave internal state for the singlet.  In momentum space s-wave and
d-wave Hamiltonian have the following forms:
\begin{equation}\label{Hsk}\begin{split}
H_s(\textbf{k})=\sum_{\textbf{k}} & \epsilon_\textbf{k}
c^\dag_\textbf{k} c_\textbf{k}+\sum_\textbf{k} \mu_f f^\dag_\textbf{k} f_\textbf{k}\\
&+V\sum_\textbf{k} (\cos(k_x)+\cos(k_y))(f^\dag_\textbf{k}
c_\textbf{k}+c^\dag_\textbf{k} f_\textbf{k})
\end{split}\end{equation}
\begin{equation}\label{Hdk}\begin{split}
H_d(k)=\sum_{k} & \epsilon_\textbf{k} c^\dag_\textbf{k} c_\textbf{k}+\sum_\textbf{k} \mu_f f^\dag_\textbf{k} f_\textbf{k}\\
&+V\sum_\textbf{k}  (\cos(k_x)-\cos(k_y))(f^\dag_\textbf{k}
c_\textbf{k}+c^\dag_\textbf{k} f_\textbf{k})
\end{split}\end{equation}
With this quadratic mean-field Hamiltonian, we can also derive the
spectrum and the mean-field parameter $V$ self-consistently:
\begin{equation}\label{specs}\begin{split}
E^s_\pm({\textbf{k}})=&\frac{\epsilon_\textbf{k}+\mu_f}{2}\pm\\
&
\sqrt{(\frac{\epsilon_\textbf{k}-\mu_f}{2})^2+V^2(\cos(k_x)+\cos(k_y))^2}
\end{split}
\end{equation}
\begin{equation}\label{specd}\begin{split}
E^d_\pm({\textbf{k}})=&\frac{\epsilon_\textbf{k}+\mu_f}{2} \pm\\
&\sqrt{(\frac{\epsilon_\textbf{k}-\mu_f}{2})^2+V^2(\cos(k_x)-\cos(k_y))^2}
\end{split}
\end{equation}
The ground state is obtained by filling all states upto the Fermi
level. As before the Fermi energy is fixed by requiring that there
are $1+n_c$ fermions in the ground state per site. Note that in both
$s$ and $d$ wave cases the $+$ band lies entirely above the $-$ band
({\em i.e} $min(E_+) \geq max(E_-)$). Therefore in the ground state
only the $E_-$ levels are occupied. The ground state energy is
\begin{equation}
E^{s,d}_{gd} = \sum_k \theta(E_F-E^{s,d}_-({\textbf{k}})) E^{s,d}_-({\textbf{k}})
\end{equation}
The self-consistency equations that determine $V$ and $\mu_f$  are now readily obtained. For instance
we have
\begin{equation}\begin{split}\label{sc}
1=\frac{J_k}{2N}\sum_\textbf{k}&
\theta(E_F-E^{s,d}_-({\textbf{k}}))\times
\\ & \frac{(\cos(k_x)\pm\cos(k_y))^2}{\sqrt{(\frac{\epsilon_\textbf{k}-\mu_f}{2})^2+V^2
(\cos(k_x)\pm\cos(k_y))^2}} \end{split}\end{equation}
 In equation \ref{sc}, $+$($-$) corresponds to s-wave(d-wave).
In addition we need to impose the condition $\langle f^{\dagger}_i
f_i\rangle = 1$ and $\langle c^{\dagger}_i c_i\rangle = n_c$.

Let us first specialize to
 the half-filled case $n_c = 1$. In this case the microscopic model has a particle-hole symmetry under which
 \begin{equation}
 c_{i\alpha} \rightarrow -i\epsilon_r\sigma^y_{\alpha \beta}
  c^{\dagger}_{i\beta}
 \end{equation}
 where $\epsilon_r = (-1)^{(x+y)} = \pm 1$ on the A and B sublattices
 of the two dimensional square lattice. As the total number of fermions per site $1+n_c = 2$ in this case, all the
$E_-$ levels are filled.  At the level of the
 approximate mean field Hamiltonians Eqn. \ref{Hammf}, under the particle-hole
 transformation
 \begin{eqnarray}
f_{i\alpha} & \rightarrow &  -i\epsilon_r\sigma^y_{\alpha \beta} f^{\dagger}_{i\beta} \\
 \mu_f & \rightarrow & -\mu_f \\
 V & \rightarrow V
 \end{eqnarray}
 Thus a particle-hole symmetric mean field state (which we assume)
 requires $\mu_f = 0$.

\begin{figure}
\includegraphics[width=4cm]{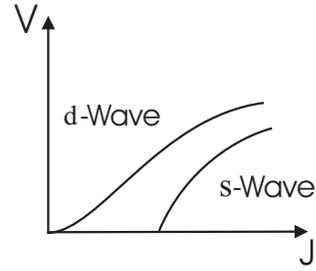}
\caption{V versus J for s-wave and d-wave self consistency
equation}\label{vjj}
\end{figure}

 With this in hand we proceed to compare the ground state energy of
 the s-wave and d-wave mean field Hamiltonians in half-filled case.
First consider self-consistency equation for s-wave case:
\begin{equation}\label{scs}
\frac{1}{J_k}=\frac{1}{2N}\sum_\textbf{k}
\frac{(\cos(k_x)+\cos(k_y))^2}{\sqrt{(\frac{\epsilon_\textbf{k}}{2})^2+V^2
(\cos(k_x)+\cos(k_y))^2}}
\end{equation}
It is obvious that the right hand side of equation \ref{scs} is
monotonically decreasing function of $|V|$. So it has its maximum
value for $V=0$:
\begin{equation}
\frac{1}{J_k}=\frac{1}{2N \sqrt{2t} }\sum_\textbf{k}
|\cos(k_x)+\cos(k_y)|
\end{equation}
The right hand-site of the above equation is finite. So there is a
maximum value of $1/J_k$ for which we can find a solution for $V$ in
equation \ref{scs}. On the other hand the self consistency equation
for d-wave at $\mu_f=0$ takes the form:
\begin{equation}\label{sccd}
\frac{1}{J_k}=\frac{1}{2N}\sum_\textbf{k}
\frac{(\cos(k_x)-\cos(k_y))^2}{\sqrt{(\frac{\epsilon_\textbf{k}}{2})^2+V^2
(\cos(k_x)-\cos(k_y))^2}}
\end{equation}
Like s-wave, the right hand site of equation \label{scd} is maximum
for $V=0$. But in this case the right hand side of equation
\ref{sccd} is divergent as $V\rightarrow 0$. So for d-wave, the
self-consistency solution exist for any value of $J_k$. A schematic
graph of $V$ versus $J_k$ for s-wave and d-wave is plotted in figure
\ref{vjj}.

 As it is also clear from the figure, there is at least a
region of small $J_k$ for which $V$ for d-wave is larger than $V$
for s-wave (in fact we can have a region with finite $V$ for d-wave
but s-wave $V$ as small as we want). Now if we look at the lower
band spectrum (eqn. \ref{specs} and \ref{specd}) this proves that
there is at least a region for small $J_k$ where d-wave ground state
energy is lower than s-wave.

Upon moving away from half-filling, we expect that the value of
ground state energy starts changing continuously; so we still have
at least a region of small $J_k$ and small doping, for which d-wave
mean-field Hamiltonian is a better approximation than s-wave. We
have confirmed this by a direct numerical solution of the
self-consistency equations.

\section{Properties of \lowercase{d}-wave Kondo liquid}
\label{dpro}
\begin{figure}[htp]
\includegraphics[width=5cm]{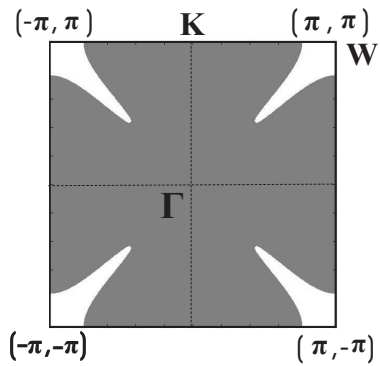}
\caption{Fermi surface in the first Brillouin zone. Occupied region
is plotted in gray.}\label{fermii}
\end{figure}

\begin{figure}[htp]
\includegraphics[width=6.2cm]{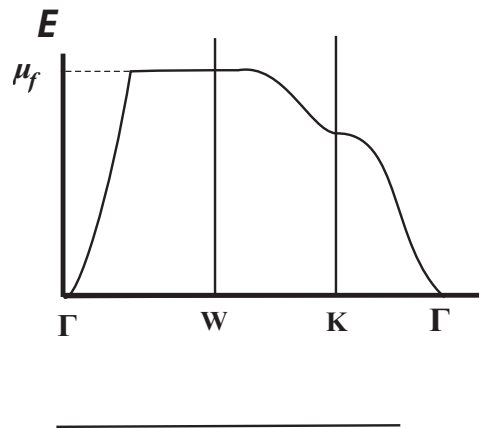}
\caption{Spectrum along the lines
$(0,0)\rightarrow(\pi,\pi)\rightarrow (0,\pi) \rightarrow
(0,0)$}\label{specc}
\end{figure}

We now
study the properties of the d-wave mean-field state away from half-filling.
 We will see that d-wave singlet formation provides a natural route
toward very anisotropic properties over the Fermi surface. Before
getting into the details of these properties, we will examine some
aspects of d-wave dispersion in equation \ref{specd}. A schematic
graph of Fermi surface for finite doping in the first Brillouin
zone, as well as dispersion along a specific path, are given in
figures \ref{fermii} and \ref{specc} respectively. One important
feature of the spectrum in figure \ref{specc} (which is also easily
proved analytically in appendix \ref{maxe}) is that the maximum
value of energy in the Brillouin zone is $\mu_f$, that is the energy
of the points along diagonal direction, where
$\mu_f<\epsilon_\textbf{k}$. This shows that Fermi surface crosses
the diagonal direction at a point where $E^d_\textbf{k}<\mu_f$
(since $E_f$ has to be less than $\mu_f$). In diagonal direction the
spectrum is
$E^d_\textbf{k}=\frac{\epsilon_\textbf{k}+\mu_f}{2}-|\frac{\epsilon_\textbf{k}-\mu_f}{2}|$
which can be written as:
\begin{equation}\nonumber
 \left\{
\begin{array}{ccc}
E^d_\textbf{k}=\epsilon_\textbf{k} & for \ \ \epsilon_\textbf{k}<\mu_f   \\
E^d_\textbf{k}=\mu_f & for \ \ \epsilon_\textbf{k}>\mu_f \end{array}
\right.
\end{equation}
combining this with the condition $E^d_\textbf{k}<\mu_f$ shows that
Fermi surface should cross the diagonal direction at a point where
the quasiparticles energy is equal to $\epsilon_\textbf{k}$. This
result is in fact the basic foundation for the interesting
properties that will be studied in the next sections i.e. over the
Fermi surface, in diagonal direction, quasi-particles are c-electron
type; the f-electron properties appear, as we go away from the
diagonal direction. In the following section we will present the
result of numerical calculation of properties over Fermi surface.

The self consistency equation for parameter $V$ could also be
studied analytically (appendix \ref{selfconsd}). We see that for
d-wave Kondo model, similar to the on site s-wave Kondo, we have $V
\propto e^{-\frac{C}{J_k}}$. Density of states at Fermi energy is
also studied analytically (appendix \ref{density}), it again shows
exponential dependence on coupling constant:
\begin{equation}
 \rho(E_f) \propto
V^{-2} \propto e^{\frac{2C}{J_k}}
\end{equation}
These results were also checked and confirmed numerically.

In the following sections we present the effective mass and
quasi-particle residue on fermi surface, as a function of angle
(from the center of Brillouin zone). We will see that these
properties are very anisotropic as we move along the Fermi surface.

\subsection{Effective mass}
\begin{figure}[htp]
\includegraphics[width=6.2cm]{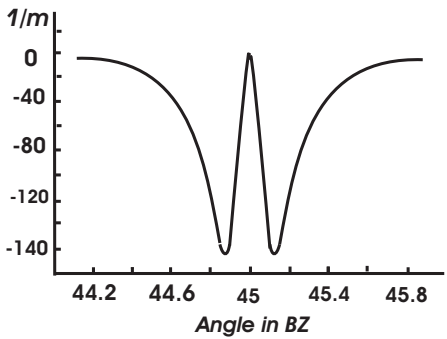}
\caption{Second derivative of energy respect to $k_\perp$ in
direction perpendicular to fermi surface.}\label{de}
\end{figure}

The effective mass of quasi-particles is defined through second
derivative of energy with respect to $k_\perp$ which is the momentum
in direction perpendicular to Fermi surface:
\begin{equation}
\frac{1}{m}=\frac{\partial^2 E^d_\textbf{k}}{\partial^2 k_\perp}
\end{equation}
A closed form for the effective mass could be derived. In figure
\ref{de} we have plotted the inverse effective mass over Fermi
surface near diagonal direction. As we expected, along diagonal
direction the effective mass matches the electron effective mass. As
we go far away from the diagonal direction, effective mass becomes
very large. This is because away from the diagonal direction, the
quasiparticle is essentially an $f$-fermion with some weak admixture
with the $c$-electron. Between these two limits we see the strange
anisotropic behavior where second derivative of energy goes from
electron type, positive value to large negative value. A comparison
with quasi-particle residue plot (figure \ref{quas}) shows that this
behavior occurs where the quasi-particles are a complete mixture of
$c$-electron and $f$-fermion.

\begin{figure}[htp]
\includegraphics[width=7.0cm]{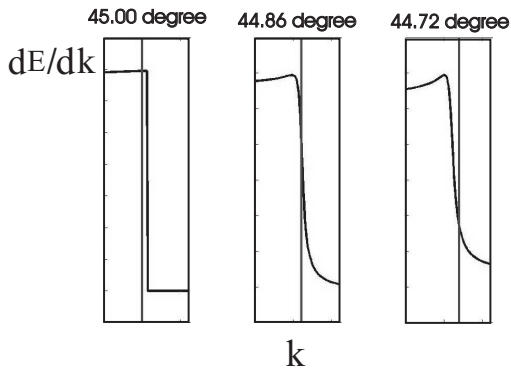}
\caption{First derivative of the energy with respect to $k_\perp$.
Vertical axis, in each plot, is $\frac{\partial E^-_k}{\partial
k_\perp}\ $ and $k_\perp$ is the momentum in direction perpendicular
to the Fermi surface at the point defined by the angle on the top.
Vertical line shows the position of the Fermi surface}\label{fd}
\end{figure}

This weird behavior could be traced by looking closer at the
spectrum in the Brillouin zone. First consider diagonal direction
(figure \ref{fd}). We see that moving along this line we go from
region where $E^-_\textbf{k}=\epsilon_\textbf{k}$ to region where
$E^-_\textbf{k}=\mu_f$ (see figure \ref{specc}). So if we plot first
derivative of energy with respect to $k_\perp$ (which is diagonal
direction for points along diagonal direction), as seen in figure
\ref{fd}, there is a jump from a finite value to zero at the point
where $\epsilon_\textbf{k}=\mu_f$. In this direction, as shown
before, at the Fermi surface $E^-_\textbf{k}=\epsilon_\textbf{k}$ so
there is a finite value for second derivative. As we move away from
diagonal direction, the jump softens slightly (because
$V_\textbf{k}$ moves slightly from zero). But still first derivative
have a large decrease in a small interval and so the second
derivative is large negative number; interestingly, Fermi surface
does cross this region at some points. This behavior is very
strange. In fact we see points at which the effective mass is much
smaller than free electron mass.

\subsection{Quasi-particle residue}
 The electron-electron green function is given by\cite{mahan}:
\begin{equation}\label{grn}
\textit{G}(k,i\omega_\nu)=\int_0^\beta d\tau
e^{i\omega_\nu\tau}\langle T_\tau c(k,\tau) c^\dag(k,0)\rangle
\end{equation}
Suppose $\gamma_\textbf{k}^+$ and $\gamma_\textbf{k}^-$ are
annihilation operators for quasi-particles in upper and lower bands
respectively, in terms of which the Hamiltonian is diagonal. The
c-electron annihilation operators could be written as:
\begin{equation}\label{qud}\begin{split}
c_\textbf{k}&=u_\textbf{k}\ \gamma_\textbf{k}^+ + v_\textbf{k}\
\gamma_\textbf{k}^- \\
v_\textbf{k}^2&=\frac{(E^+_\textbf{k}-\epsilon_\textbf{k})^2}{(E^+_\textbf{k}-\epsilon_\textbf{k})^2+V_\textbf{k}^2}
 \\
u_\textbf{k}^2&=\frac{V_\textbf{k}^2}{(E^+_\textbf{k}-\epsilon_\textbf{k})^2+V_\textbf{k}^2} \\
u_\textbf{k}&=-\frac{V_\textbf{k}
v_\textbf{k}}{E^+_\textbf{k}-\epsilon_\textbf{k}}
\end{split}
\end{equation}

\begin{figure}[htp]
\includegraphics[width=6.2cm]{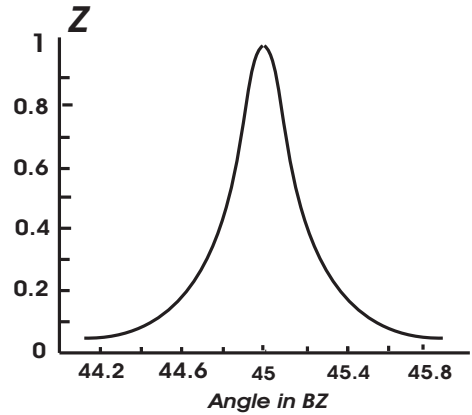}
\caption{Quasi-particle residue on the fermi surface}\label{quas}
\end{figure}

Using this form (and the fact that Hamiltonian is diagonal in terms
of $\gamma^\pm_\textbf{k}$ operators):
\begin{equation}\nonumber
\langle c(\textbf{k},\tau)
c^\dag(\textbf{k},0)\rangle=u_\textbf{k}^2\langle
\gamma^+_\textbf{k}(\tau) \gamma^{+\dag}_\textbf{k}(0) \rangle +
v_\textbf{k}^2\langle \gamma^-_\textbf{k}(\tau)
\gamma^{-\dag}_\textbf{k}(0) \rangle
\end{equation}
Rewriting the Hamiltonian in term of $\gamma$ operators
($H=\sum_\textbf{k}(\gamma^{+\dag}_\textbf{k}\gamma^+_\textbf{k}+\gamma^{-\dag}_\textbf{k}\gamma^-_\textbf{k})$)
we get:
\begin{equation}\nonumber
 \langle \gamma^\pm_\textbf{k}(\tau) \gamma^{\pm\dag}_\textbf{k}(0)
 \rangle=e^{-E^\pm_\textbf{k}\tau}
\end{equation}
 Putting these into equation \ref{grn} gives:
\begin{equation}
\textit{G}(k,i\omega_\nu)=\frac{u_\textbf{k}^2}{E_\textbf{k}^+-i\omega_{\nu}}+\frac{v_\textbf{k}^2}{E_\textbf{k}^--i\omega_{\nu}}
\end{equation}
By analytically continuing this to real frequencies and taking the
imaginary part we get the spectral function\cite{mahan}:
\begin{equation}\label{spf}
A(k,\omega)=u_\textbf{k}^2
\delta(\omega-E^+_\textbf{k})+v_\textbf{k}^2
\delta(\omega-E^-_\textbf{k})
\end{equation}

This consists of two peaks and the weight under the low energy peak
will give us the quasiparticle residue (Z). We have plotted Z for
the points near diagonal direction in figure \ref{quas}. Again as we
expect, the quasiparticle residue near diagonal direction is of
order one (electron type excitations) and gets very small away from
diagonal direction.

\section{\lowercase{d}-wave Kondo liquid in a Kondo-Heisenberg model: Fermi liquid with spinons}
\label{dklh}
We now study the properties of the d-wave Kondo
liquid state in a model which allows for explicit Heisenberg
exchange interactions between the local moments. Remarkably we
will show that the quasiparticle residue vanishes at isolated
points of the Fermi surface in such a state. The excitation at
such points is a free neutral fermionic spinon. In a subsequent
Section, we show that this spinon survives even when fluctuations
beyond the mean field are included. Thus this d-wave Kondo liquid
is a Fermi liquid state that supports spinons at isolated Fermi
points. Specifically we consider the model
\begin{equation}\label{dhkondo}
H=\sum_{\langle ij \rangle} t_{ij} c^\dag_i c_j+J_K \sum_{\langle ij
\rangle} \vec{S}_i.\vec{s}_j+J_H \sum_{\langle ij
\rangle}\vec{S}_i.\vec{S}_j
\end{equation}

We proceed as before using the d-wave mean-field approximation to
treat $\vec{S}_i.\vec{s}_j$ term; but here, we have another
interacting term, which is the direct Heisenberg exchange between
the local moments. Expressing this in terms of the $f$-fermions
gives rise to a four fermion term. We treat this in mean field
theory as well. While a number of different mean field decouplings
are possible, we focus here on one in the particle-hole channel
which endows the $f$-fermions with a uniform non-zero hopping
$\chi$. This is in turn determined self-consistently through the
equation
\begin{equation}
\chi=J_H \langle f^\dagger_i f_j \rangle
\end{equation}
Using this, we get the following mean-field Hamiltonian:
\begin{equation}\label{Hdkh}\begin{split}
H_H(k)=\sum_{k} & \epsilon_\textbf{k} c^\dag_\textbf{k} c_\textbf{k}+\sum_\textbf{k} \epsilon_{f\textbf{k}} f^\dag_\textbf{k} f_\textbf{k}\\
&+V\sum_\textbf{k}  (\cos(k_x)-\cos(k_y))(f^\dag_\textbf{k}
c_\textbf{k}+c^\dag_\textbf{k} f_\textbf{k})
\end{split}\end{equation}
where $\epsilon_{f\textbf{k}}=\mu_f-\chi\ (\cos(k_x)+\cos(k_y))$.
With this quadratic Hamiltonian, we can again get the spectrum:

\begin{equation}\label{spech}\begin{split}
E^H_\pm({\textbf{k}})=&\frac{\epsilon_\textbf{k}+\epsilon_{f\textbf{k}}}{2} \pm\\
&\sqrt{(\frac{\epsilon_\textbf{k}-\epsilon_{f\textbf{k}}}{2})^2+V^2(\cos(k_x)-\cos(k_y))^2}
\end{split}
\end{equation}

Now let us have a closer look at this Hamiltonian and spectrum. When
$V=0$, local moments and conduction electrons are not hybridized and
we have two separated Fermi surfaces. The electron Fermi surface is
identified by spectrum $\epsilon_\textbf{k}$ and is small. Spinon
Fermi surface, contains one moment per site and covers half the
Brillouin zone\cite{senvoj}.

Now assume turning on non-zero $V$. If $V$ is small enough, both
bands intersect the Fermi energy.  The resulting Fermi surface then
consists of two sheets (each identified with one of the bands).
Consider the quasi-particle residue on each band. From Eqn.
\ref{spf}, it is clear that on the Fermi surface of the $E^H_-$
band, quasi-particle residue is given by $v_\textbf{k}^2$ while on
the other sheet (associated with $E^H_+$) it is given by
$u_\textbf{k}^2$. $v_\textbf{k}$ and $u_\textbf{k}$ are defined in
Eqn. \ref{qud} and satisfy
\begin{eqnarray}\label{uv}
u_\textbf{k}&=&-\frac{V(\cos(k_x)-\cos(k_y))}{E^H_+(\textbf{k})-\epsilon_\textbf{k}}\ v_\textbf{k}\nonumber\\
u_\textbf{k}^2&+&v_\textbf{k}^2=1
\end{eqnarray}

Using this we see that the Fermi surface of the $-$ band has large
quasiparticle residue and thus has essentially $c$-electron
character (with weak admixture to $f$-fermions). On the other hand
the $+$ Fermi surface has small quasiparticle residue and has
essentially $f$-fermion character with weak admixture to
$c$-electrons.  Following reference [\onlinecite{senvoj}] we name
the $-$ and $+$ Fermi surfaces as cold and hot surface
respectively. Remarkably the quasiparticle residue on the hot
Fermi surface vanishes at four isolated points (which are along
the diagonal directions). At these four points the excitation is a
pure $f$-fermion with {\em no} admixture to the $c$-electron. Thus
at these isolated points the excitation is a neutral fermionic
spinon even though spinons do not exist elsewhere on the Fermi
surface.

\begin{figure}[htp]
\includegraphics[width=6cm]{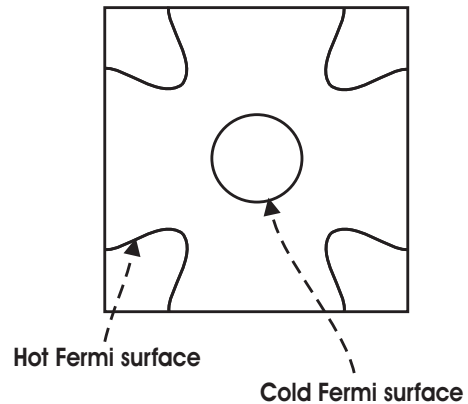}
\caption{Fermi surface of the d-wave Kondo liquid in the Kondo-Heisenberg model }\label{khfig}
\end{figure}

\section{Quantum Hall Kondo insulators}
\label{kiqh} In this Section we describe an interesting Kondo
insulating state that is possible if the Kondo singlet has
nontrivial internal angular momentum. Consider the generalized Kondo
Hamiltonian again (\ref{dkondo}) and assume $J_{ij}$ is non-zero for
nearest ($J_1$) and next nearest neighbors ($J_2$). Now, as before,
we decouple the interacting term using an auxiliary field and
consider the saddle point solution with nearest neighbor
$V(\tau,i,j)$  the same as d-wave, and next nearest neighbor equal to
$iV_2$ along one diagonal direction and $-iV_2$ along the other.
Clearly this coresponds to a Kondo singlet with internal angular
momentum $d_{x^2-y^2} + id_{xy}$, and describes a mean field state
that spontaneously breaks time reversal symmetry. This leads to the
following mean-field Hamiltonian:
\begin{equation}\begin{split}\label{did}
H_{d+id}(\textbf{k})=\sum_{\textbf{k}} & \epsilon_\textbf{k} c^\dag_\textbf{k} c_\textbf{k}+\sum_\textbf{k} \mu_f f^\dag_\textbf{k} f_\textbf{k}\\
&+V_1\sum_\textbf{k}  (\cos(k_x)-\cos(k_y))(f^\dag_\textbf{k}
c_\textbf{k}+c^\dag_\textbf{k} f_\textbf{k})\\ &+ V_2
\sum_\textbf{k} i\sin(k_x)\sin(k_y) (c^\dag_\textbf{k}
f_\textbf{k}-f^\dag_\textbf{k} c_\textbf{k})
\end{split}\end{equation}
We now specialize to a half-filled conduction band $n_c = 1$.
Diagonalizing this Hamiltonian, we readily see that the ground state
in this case is a Kondo insulator with a gap. However we now show
that the broken time reversal symmetry leads to a non-vanishing
quantized electrical Hall conductivity. Thus this state provides an
interesting example of a quantum Hall effect in a local moment
system driven by the Kondo effect.
 To expose this physics it is convenient to rewrite the Hamiltonian in terms of a two-component fermion operator
$\psi_\textbf{k}$:
\begin{equation}\nonumber
 \psi_\textbf{k}= \left(
\begin{array}{ccc}
c_\textbf{k} \\
f_\textbf{k} \end{array} \right)
\end{equation}
We have
\begin{equation}\label{hamil}
H=\sum_\textbf{k} \psi_\textbf{k}^\dag \left(
\frac{\epsilon_\textbf{k}+\mu_f}{2}+\vec{m}(\textbf{k}).\vec{\tau}
\right)\psi_\textbf{k}
\end{equation}
Here $\vec \tau$ are Pauli matrices operating on the two components of $\psi_\textbf{k}$ and
$\vec{m}_\textbf{k}$ is a 3-dimensional vector defined in the two dimensional Brillouin zone:
\begin{equation}\label{mf}\begin{split}
\vec{m}_\textbf{k}=&\ \ V_1(\cos(k_x)-\cos(k_y))\hat{x}\\ &+ V_2
\sin(k_x)\sin(k_y)\hat{y}\\ &+\frac{\epsilon_\textbf{k}-\mu_f}{2}\
\hat{z}
\end{split}\end{equation}
Note that if $|\vec{m}|\neq 0$ over the Brillouin zone,
$\vec{m}/|\vec{m}|$ is topologically a mapping from a torus
(Brillouin zone) to the unite sphere. To calculate the Hall
conductivity, we need the current operator. Note that in this mean
field Hamiltonian, charge conservation symmetry is realized as
invariance under $\psi_\textbf{k} \rightarrow e^{i\alpha}
\psi_\textbf{k}$. Thus though the $f$-fermions start off as neutral,
the mean field condensation of $V$ has endowed them with physical
charge. This observation then leads to the current operator:
\begin{equation}\label{curr}
J_\mu(\textbf{k})=\frac{\partial}{k_\mu}H
\end{equation}
We can now calculate $\sigma_{xy}$ using the Kubo
formula\cite{kubo}. The details of the calculation are in Appendix
\ref{kb}. We get
\begin{equation}\label{topol}
\sigma_{xy}=\frac{e^2}{4\pi h}\int d^2k \frac{\vec{m}.(\partial_x
\vec{m} \times \partial_y\vec{m})}{|\vec{m}|^3}
\end{equation}
 The value of $\sigma_{xy}$ is then invariant under smooth change of
 $\vec{m}$. It is thus a topological invariant of the kind considered previously by Volovik\cite{volovik}.
We can now change $\vec{m}$ smoothly to a mapping for which
integration \ref{topol} could be calculated easily and gives twice the surface area
of a unit sphere so that
\begin{equation}
\sigma_{xy}=\frac{2 e^2}{h}
\end{equation}
This is also confirmed with direct numerical integration. The
essential physics is also simply illustrated  by the following
argument. As the relevant integral is a topological invariant we
first imagine a smooth deformation to make $V_2$ infinitesimally
small. In the limit when $V_2 = 0$ the gap closes and
 $|\vec{m}(\textbf{k})|$ has four zeroes in the Brillouin zone on the
 diagonal directions where $\epsilon_\textbf{k}=\mu_f$. At low energies, the physics will be dominated by modes near
 these nodes. On turning on a small value of
 $V_2$, we expect that the universal physics is still correctly captured, in an approximation that is legitimate for modes
 near the nodes. So we
expand $\vec{m}(\textbf{k})$ near these nodes and take the integral
  over $k_x$ and $k_y$ from $-\infty$ to $\infty$. After the
 expansion, we get (for the node at $(k_p, k_p)$ in the first quadrant of the Brilloun zone):
 \begin{equation}\label{dm}\begin{split}
\vec{m}_q=&\ \ V_1 \sin(k_p)\frac{q_x-q_y}{2}\hat{x}\\ &+ V_2 \sin^2(k_p)\hat{y}\\
&+t\sin(k_p)\frac{q_x+q_y}{2}\ \hat{z}
\end{split}\end{equation}
 Here $q_i=k_i-k_p$ and $\cos(k_p)=\frac{\mu_f}{4t}$. After renaming
$q_x-q_y$ as $k_x$, $q_x+q_y$ as $k_y$, $V_1 \sin(k_p)$ as
$\textit{v}_x$, $t \sin{k_p}/2$ as $\textit{v}_y$ and $V_2
\sin(k_p)^2$ as $\Delta$, along with $90^\circ$  rotation of
$\vec{m}$ around $x$ axis, we get the following simple form:
\begin{equation}
\vec{m}(\textbf{k})=\textit{v}_x k_x \hat{x}+\textit{v}_y k_y
\hat{y}+\Delta \hat{z}
\end{equation}
 From this we trivially see that
$
\hat{m}(\textbf{k})=\frac{\vec{m}(\textbf{k})}{|\vec{m}(\textbf{k})|}$
points along $\hat{z}$ at $k_x=k_y=0$ and points along radial
direction in $k_x,k_y$ plane as $\sqrt{k_x^2+k_y^2}\rightarrow
\infty$. So this vector covers half the sphere and contributes $2
\pi$ to the integral in Eqn. \ref{topol}. Exactly the same
contribution arises from the other nodes and so we get $4\times
2\pi= 8\pi$. Putting this in \ref{topol} gives the result quoted
above for the electrical Hall conductivity.

Thus the $d_{x^2-y^2}+ id_{xy}$ Kondo insulator has a quantized
electrical Hall conductivity. A similar result - a spin and thermal
quantum Hall effect - was established for two dimensional
$d_{x^2-y^2}+ id_{xy}$ {\em superconductors} in Ref.
\onlinecite{sqh}.

\section{Beyond mean field theory}
Through out this paper we have treated the Kondo and other
interactions within the slave boson mean field theory. Now we
consider the role of fluctuations beyond this mean field theory. It
has been known for a long time that the important fluctuations in
such mean field theories are gauge fluctuations\cite{bas}.
Specifically the redundant representation of the local moment spin
$\vec S_i$ in terms of $f$-fermions leads to invariance of the
resulting theory under local $U(1)$ gauge
transformations\footnote{Actually the redundancy leads to invariance
under $SU(2)$ gauge transformations but our main point can be made
by focusing just on the $U(1)$ subgroup.} where
\begin{equation}
f_i \rightarrow e^{i\alpha_i} f_i
\end{equation}
The hybridization field in turn transforms as
\begin{equation}
V_{ij} \rightarrow e^{-i\alpha_i} V_{ij}
\end{equation}
Note that the hybridization field also carries physical electric
charge equal to the electron charge. As in the usual on-site slave
boson theory, the mean field state is a Higgs phase where the
hybridization field is condensed. Consequently the gauge
fluctuations acquire a gap and can be integrated out. The Higgs
condensate also implies that the $U(1)$ gauge charge of a localized
$f$-fermion  will be screened to produce a gauge neutral object that
carries physical electric charge $1$. Thus exactly as in the usual
on-site slave boson theory, the gauge fluctuations will play an
innocuous role in these mean field states, and the mean field theory
provides a correct qualitative description of the universal features
of these phases. These considerations apply even to the $d$-wave
Kondo liquid of Section \ref{dklh} where the hybridization amplitude
vanishes at four points of the `hot' Fermi surface. Thus the
vanishing quasiparticle residue at those four points (and the
associated spinon excitations) survive beyond the mean field
approximation.

\section{Discussion}
\label{disc} The considerations of this paper may be useful in
future theoretical work on a number of different correlated electron
materials. The example of the cuprate materials shows that
correlation effects may not uniformly affect all regions of the
Fermi surface. The result is strong correlation induced anisotropy
along the Fermi surface. Theoretical approaches to addressing such
effects are hampered by the difficulty that correlation effects are
easiest to handle in real space and not in momentum space. In this
paper in the specific case of Kondo lattices, we have shown how to
incorporate momentum space information into the Kondo singlet
formation that determines the fate of the local moments at low
temperature. We explored the properties of metallic `Fermi liquid'
states  driven by Kondo singlet formation in a channel with non-zero
internal angular momentum. Through out the paper we focused on two
dimensional systems, though extension of our results to three
dimensions is straightforward. We showed that such metallic states
naturally have strong anisotropy of the quasiparticle effective mass
and residue on moving around the Fermi surface. In some cases the
quasiparticle residue even vanishes at four isolated points on the
Fermi surface. The excitations at such points may be thought of as
neutral spin-$1/2$ spinons that occur without any residual `gauge'
interactions. Thus these states provide interesting examples where
strongly anisotropic quasiparticle residues are naturally built into
the symmetry of the states. We also studied two dimensional Kondo
{\em insulators} driven by Kondo singlet formation with complex
internal angular momentum and showed that they have a quantized
non-trivial electrical Hall conductivity. Such Kondo insulators thus
present an interesting situation where a quantum Hall effect occurs
due to the Kondo effect. Exploiting the ideas of this paper to
develop techniques for thinking about the angle dependence of
correlation effects in momentum space is an interesting challenge
for the future.

\section*{Acknowledgments}
We thank P.A. Lee, A.J. Millis, and Subir Sachdev for useful discussions.
This work was funded by NSF Grant No. DMR-0308945.
TS also acknowledges
funding from the Alfred P. Sloan Foundation, an award from the The Research Corporation, and a DAE-SRC Outstanding Investigator Award in India.

\appendix
\section{Maximum energy in Brillouin  zone}\label{maxe}
As derived before, lower band energy at  each point of Brillouin
zone (BZ) is given by
\begin{equation}\label{aspc}
E^-_{\textbf{k}}=\frac{\epsilon_\textbf{k}+\mu_f}{2} -
\sqrt{(\frac{\epsilon_\textbf{k}-\mu_f}{2})^2+V_\textbf{k}^2}
\end{equation} where
$V_\textbf{k}=V(\cos(k_x)-\cos(k_y))$. So $V_\textbf{k}=0$ only
along diagonal directions. From \ref{aspc}, we see for all points in
BZ:
\begin{equation}\label{aineq}
E^-_{\textbf{k}}\leq\frac{\epsilon_\textbf{k}+\mu_f}{2} -
\sqrt{(\frac{\epsilon_\textbf{k}-\mu_f}{2})^2}
\end{equation}
Right hand side of the above equation is equal to $\mu_f$ for
$\mu_f<\epsilon_\textbf{k}$ and $\epsilon_\textbf{k}$ for
$\epsilon_\textbf{k}<\mu_f$. Putting these together we get:
\begin{equation}
 \left\{
\begin{array}{ccc}
E^-_\textbf{k}\leq\epsilon_\textbf{k} & for \ \ \epsilon_\textbf{k}\leq\mu_f   \\
E^-_\textbf{k}\leq\mu_f & for \ \ \epsilon_\textbf{k}>\mu_f
\end{array} \right.
\end{equation}
You see that  all points in BZ, obviously, satisfy one the two
conditions ($\epsilon_\textbf{k}\leq\mu_f$ or
$\epsilon_\textbf{k}>\mu_f$). So we get for any point in the BZ:
\begin{equation}\label{amr}
E^-_\textbf{k}\leq\mu_f
\end{equation}
A closer look at equations \ref{aspc} and \ref{aineq}, shows that
equality in equation \ref{amr}, could be only for the points along
diagonal direction with $\epsilon_\textbf{k}>\mu_f$. Such points
cover a region with zero volume in the BZ. So that for any finite
doping $E_f<\mu_f$ (see section \ref{dpro}).

\section{Self consistency equation}\label{selfconsd}
Analytic treatment of self consistency relation is easier and much
more clear at zero doping ($\mu_f=0$). So we start with this case
and later, we can argue if our result might be modified at finite,
but small, doping. In continuum limit, self consistency relation (at
zero doping) have the form:
\begin{equation}\label{sci}
\frac{1}{J_k}=\frac{1}{2n}\int \frac{d^2k}{(2\pi)^2}
\frac{(\cos(k_x)-\cos(k_y))^2}{\sqrt{(\frac{\epsilon_\textbf{k}}{2})^2+V^2(\cos(k_x)-\cos(k_y))^2}}
\end{equation}
where $n$ is density of lattice sites. We saw before that, this
equation for $V$ has a solution, no matter how small $J_k$ is (see
section \ref{lecomp}). This was because, the integral is divergent
at $V=0$. The divergence comes from the region where
$\epsilon_\textbf{k}$ is small. So to study the behavior of this
integral, it is enough to look at the points where
$\epsilon_\textbf{k}$ is small. Because of the symmetry of BZ with
respect to $90^\circ$ rotation, it is enough to consider region with
$k_x,k_y\geq 0$. The point with $\epsilon_\textbf{k}=0$ in this
region, are points with $k_x=\pi-k_y$. We want to look around such
points so if we change the variables to $k_y\rightarrow k_y$ and
$k_x\rightarrow q=k_y+k_x-\pi$, we just want to look at region with
small $q$. Putting these variables in to in to \ref{sci} and
expanding up to first non-zero order in $q$, we get:
\begin{equation}\label{scis}\begin{split}
\frac{1}{J_k}=\frac{1}{2n}\int & \frac{dk_y dq}{(2\pi)^2}
\\ & \frac{(\sin(k_y)q+2\cos(k_y))^2}{\sqrt{(t\sin(k_y)q)^2+V^2(\sin(k_y)q+2\cos(k_y))^2}}
\end{split}
\end{equation}
As discussed before, we like to study the singular behavior around
$q=0$ so we can ignore the $q$ dependence in non-singular terms. The
relation simplifies to:
\begin{equation}\label{sciss}
\frac{1}{J_k}=\frac{1}{2n}\int\frac{dk_y
dq}{(2\pi)^2}\frac{(2\cos(k_y))^2}{\sqrt{(t\sin(k_y)q)^2+V^2(2\cos(k_y))^2}}
\end{equation}
Now we first perform the integral over $k_y$. Note that the limits
for this is $q$ dependent (i.e. $0+O(q^2)$ and $\pi+O(q^2)$) but
these also could be ignored since they have no effect on singular
behavior:
\begin{equation}\label{scif}
\frac{1}{J_k}=\frac{1}{2n}\int\frac{dq}{2\pi}\int_0^\pi\frac{dk_y
}{(2\pi)}\frac{(2\cos(k_y))^2}{\sqrt{(t\sin(k_y)q)^2+V^2(2\cos(k_y))^2}}
\end{equation}
Integral over $k_y$ could be perform exactly which gives:
\begin{equation}
\frac{2}{|q|}\ _2F_1(\frac{1}{2},\frac{3}{2};2;1-\frac{V^2}{q^2})
\end{equation}
here, $\-_2F_1$ is Hypergeometric Function. We are in interested in
small $q$ behavior so $1-\frac{V^2}{q^2}\approx-\frac{V^2}{q^2}$.
The expression could be simplified using the identity\cite{hyp}:
\begin{equation}\nonumber
\ _2F_1(a,b;c;z)=(1-z)^{-a}\ _2F_1(a,c-b;c;\frac{z}{z-1})
\end{equation}
Using this for the result of integral we get:
\begin{equation}
\frac{1}{J_k}\approx\frac{1}{2n}\int\frac{dq}{2\pi}\frac{2}{\sqrt{q^2+V^2}}\
_2F_1(\frac{1}{2},\frac{1}{2};2;1)
\end{equation}
where we used the fact
$\frac{-\frac{V^2}{q^2}}{-\frac{V^2}{q^2}-1}\approx1$. Since
$_2F_1(\frac{1}{2},\frac{1}{2};2;1)$ is  convergent\cite{hyp} we
get:
\begin{equation}
\frac{1}{J_k}=A\int\frac{dq}{\sqrt{q^2+V^2}}
\end{equation}
where $A$ is a finite constant. Integral over $q$ is now trivial and
leads to result mentioned before (see section \ref{dpro}):
\begin{equation}\label{expo}
V \propto e^{-\frac{C}{J_k}}
\end{equation}
 After dopinping, continuum form of self consistency
 relation changes:
\begin{equation}\label{scidop}\begin{split}
\frac{1}{J_k}=\frac{1}{2n}\int\frac{d^2k}{(2\pi)^2}&\ \Theta(E_f-E^-\textbf{k})\\
&\frac{(\cos(k_x)-\cos(k_y))^2}{\sqrt{(\frac{\epsilon_\textbf{k}-\mu_f}{2})^2+V^2(\cos(k_x)-\cos(k_y))^2}}
\end{split}
\end{equation}
Again dominant contribution comes from the region where
$|\epsilon_\textbf{k}-\mu_f|$ is small. Similar expansion could be
carried out but this time the points we look at are close to
different curve (defined by $\cos(k_x)+\cos(k_y)=\frac{\mu_f}{2t}$
which is slightly away from $k_x=\pi-k_y$ for small $\mu_f$). We
don't expect (and in fact the resulting integral shows) that there
is not much change in singular behavior as $V$ goes to zero. So we
expect that behavior seen in \ref{expo} holds and in fact this is
confirmed with numerical studies.

\section{Density of states}\label{density}
To get density of states we use the general formula\cite{ashk}:
\begin{equation}
\rho(E_f)=\int \frac{d^2k}{(2\pi)^2}\  \delta(E_f-E^-_\textbf{k})
\end{equation}

Again symmetries in BZ are helpful. First of all we can do the
calculation on one patch of Fermi surface (in $k_x\leq 0$ and
$k_y\leq 0$ quarter). Also because of reflection symmetry respect to
diagonal, we can do the calculation for the points on Fermi surface
which are above the diagonal and double the result. Now in this
region we can safely (since $\epsilon_\textbf{k}$ is one to one
function of $(k_x,k_y)$) do the change of variable, and work with
$k_y$ and $\epsilon_\textbf{k}$:
\begin{equation}\label{denall}
\rho(E_f)=\int
\frac{dk_y}{(2\pi)^2}\frac{d\epsilon_\textbf{k}}{2t\sqrt{1-(\frac{\epsilon_\textbf{k}}{2t}-\cos(k_y))^2}}
\ \delta(E_f-E^-_\textbf{k})
\end{equation}
Note that
$\sqrt{1-(\frac{\epsilon_\textbf{k}}{2t}-\cos(k_y))^2}=\sin(k_x)$ is
never zero in the part of BZ we are integrating over. In terms of
new variables:
\begin{equation}
E^-_\textbf{k}=\frac{\epsilon_\textbf{k}+\mu_f}{2}-\sqrt{(\frac{\epsilon_\textbf{k}-\mu_f}{2})^2+V^2(\frac{\epsilon_\textbf{k}}{2t}-2\cos(k_y))^2}
\end{equation}
Using this we first do the integration over $\epsilon_\textbf{k}$
which leads to:
\begin{equation}\label{dendi}\begin{split}
\rho(E_f)=\int &
\frac{dk_y}{(2\pi)^2}\frac{1}{2t\sqrt{1-(\frac{\epsilon}{2t}-\cos(k_y))^2}}\\
&
\left(\frac{1}{2}-\frac{\frac{\epsilon-\mu_f}{4}+(V^2/2t)(\frac{\epsilon}{2t}-2\cos(k_y))}{\sqrt{(\frac{\epsilon-\mu_f}{2})^2+V^2(\frac{\epsilon_\textbf{k}}{2t}-2\cos(k_y))^2}}\right)^{-1}
\end{split}\end{equation}
here, $\epsilon$ is solution of equation:
\begin{equation}\nonumber
E_\textbf{k}^-=E_f
\end{equation}
for $\epsilon_\textbf{k}$, which is a simple second order equation
but we avoid presenting the answer which is not necessary for the
rest of calculation. Now to proceed we need to divide to points in
three different regions. The first region is defined by the points
where:
\begin{equation}\nonumber
(\frac{\epsilon_\textbf{k}-\mu_f}{2})^2 \sim
V^2(\frac{\epsilon_\textbf{k}}{2t}-2\cos(k_y))^2
\end{equation}
for these points contribution to \ref{dendi} is of order one.

The other regions are where:
\begin{equation}
(\frac{\epsilon_\textbf{k}-\mu_f}{2})^2\gg
V^2(\frac{\epsilon_\textbf{k}}{2t}-2\cos(k_y))^2
\end{equation}
This contains most of the points on the Fermi surface, since from
\ref{expo} we know that $V$ is small. Now we expand expression in
prentices in \ref{dendi} up to first non-zero order in $V^2$:

\begin{equation} \rho(E_f)=\int
\frac{dk_y}{(2\pi)^2}
\left(\frac{1}{2}(1-Sign(\epsilon_\textbf{k}-\mu_f))+\frac{V^2}{t^2}
f(\epsilon,k_y)\right)^{-1}
\end{equation}
here $f(\epsilon,k_y)$ is convergent function. We don't need the
detailed form of this function though. The important property of
this integral is that for points with $\epsilon_\textbf{k}<\mu_f$
(which corresponds to points near diagonal as discussed in
\ref{maxe}) contribution to integral is of order one. We expected
this since in this region quasi-particles are more free electron
like. However for points with $\epsilon_\textbf{k}>\mu_f$ (which are
away from diagonal) contribution is large and proportional to
$V^{-2}$. Putting all these together we get a large density of
state:
\begin{equation}
\rho(E_f)\propto e^\frac{2C}{J_k}
\end{equation}

\section{Kubo calculation}\label{kb}
Using equation \ref{curr} we can write $J_\mu$ as:
\begin{equation}\label{curr2}
J_\mu(\tau)=\sum_\textbf{k} \psi^\dag(\tau) (\partial_\mu
a(\textbf{k})+\partial_\mu
\vec{m(\textbf{k})}.\vec{\sigma})\psi(\tau)
\end{equation}
where $\vec{m(\textbf{k})}$ is defined in \ref{mf} and
$a(\textbf{k})=\frac{\epsilon_\textbf{k}+\mu_f}{2}$. Also
$\partial_\mu$ is the shorthand for $\frac{\partial}{\partial
k_\mu}$. Putting this form in Kubo formula\cite{kubo} we get:
\begin{equation}\begin{split}
\sigma_{\mu\nu}(\omega)=\int  d\tau &\frac{e^{-i\omega
\tau}-1}{\omega} \times \\ \sum_{k,k'}\langle &
\psi^\dag(\tau)(\partial_\mu a(\textbf{k})+\partial_\mu
\vec{m}(\textbf{k}).\vec{\sigma})\psi(\tau)\\\ \  & \psi^\dag(0)
(\partial_\nu a(k')+\partial_\nu
\vec{m}(k').\vec{\sigma})\psi(0)\rangle
\end{split}\end{equation}
Using the Hamiltonian given in \ref{hamil} we get the green function
for $\psi$ fields:
\begin{equation}\label{green}
G(i\omega,k)=\frac{1}{i\omega-a(\textbf{k})-\vec{m}(\textbf{k}).\vec{\sigma}}
\end{equation}
With this in hand, the Kubo equation reduces to the following form:
\begin{equation}\begin{split}
&\omega  \sigma_{xy}(\omega)=\sum_\textbf{k}\int \frac{dE}{2\pi}\\
&tr\left[\frac{(\partial_x a(\textbf{k})+\partial_x
\vec{m}(\textbf{k}).\vec{\sigma})(\partial_y
a(\textbf{k})+\partial_y
\vec{m}(\textbf{k}).\vec{\sigma})}{(iE-a(\textbf{k})-\vec{m}(\textbf{k}).\vec{\sigma})}\right.\times\\
&\left.\left(\frac{1}{(i(\omega+E)-a(\textbf{k})-\vec{m}(\textbf{k}).\vec{\sigma})}-\frac{1}{(iE-a(\textbf{k})-\vec{m}(\textbf{k}).\vec{\sigma})}\right)\right]
\end{split}\end{equation}
We are interested in the limit of above equation as $\omega
\rightarrow 0$. In this limit we get the following expression for
Hall conductivity:
\begin{equation}\begin{split}
\sigma_{xy}=&-i\int\frac{d^2k}{(2\pi)^2}\frac{dE}{2\pi}\\ & tr\left[
(\partial_x a(\textbf{k})+\partial_x
\vec{m}(\textbf{k}).\vec{\sigma})\frac{1}{(iE-a(\textbf{k})-\vec{m}(\textbf{k}).\vec{\sigma})^2}
\right.\\ &\left.(\partial_y a(\textbf{k})+\partial_y
\vec{m}(\textbf{k}).\vec{\sigma})\frac{1}{(iE-a(\textbf{k})-\vec{m}(\textbf{k}).\vec{\sigma})}\right]
\end{split}
\end{equation}
Expanding out the sums, we have several terms but taking the trace,
cancel some of the terms. After integration over the $E$ and
dropping the terms which are zero under the trace gives:
\begin{equation}\begin{split}
\sigma_{xy}=-i\int\frac{d^2 k}{(2\pi)^2}&\frac{1}{8
|\vec{m}(\textbf{k})|^3}\\&tr\left(\partial_x\vec{m}(\textbf{k})\right.\left.
[\vec{m}(\textbf{k}).\vec{\sigma},\partial_y
a(\textbf{k})+\partial_y \vec{m}(\textbf{k}).\vec{\sigma}]\right)
\end{split}\end{equation}
After doing some algebra on the Pauli matrices and taking the trace,
we get the relation given in \ref{topol}.
\bibliography{dkondo3}
\end{document}